\documentclass[a4paper]{article}
\usepackage[english]{babel}
\usepackage{icrc2013}

\title{Optimized next-neighbor image cleaning method for Cherenkov Telescopes}

\shorttitle{ICRC 2013 Template}

\authors{
M. Shayduk$^{1}$,
for the CTA Consortium.
}

\afiliations{
$^1$ DESY Zeuthen \\
}

\email{maxim.shayduk@desy.de}

\abstract{ 
In photo-sensor cameras of Cherenkov telescopes the light images from particle showers always contain 
the background noise induced by photons of the night sky. An image cleaning procedure is needed to reduce 
the contribution of those noise photons in further analysis stages. The conventional topological next-neighbor method 
lacks reconstruction efficiency for low light content images and image peripheries with low signal levels. 
We present here a simple optimization of the traditional next-neighbor image cleaning method that exploits the limited
time duration of shower flashes and short time-difference between neighboring image pixels. 
This method reduces greatly the noise contribution by applying dynamical cuts in the parameter space formed by signal 
amplitude and time-difference between neighboring pixels
}


\begin{document}
\maketitle

\section{Introduction}

Very high energy (VHE) ground-based $\gamma$-ray astronomy  has rapidly 
developed over the last several decades. With an introduction of the Imaging Atmospheric Cherenkov telescopes technique, pioneered by the
Whipple \cite{bib:whipple},\cite{bib:whipple2} and HEGRA \cite{bib:hegra}  collaborations it has reached a very productive state  with the currently operating experiments H.E.S.S. \cite{bib:hess},
MAGIC \cite{bib:magic} and VERITAS \cite{bib:veritas}. The next generation ground-based $\gamma$-ray  instrument - Cherenkov Telescope Array (CTA) \cite{bib:CTA}  is aiming to reach an order of magnitude higher sensitivity,
compared to currently running facilities, extending the energy range at the same time. It will consist of about 60 prime-focus Cherenkov telescopes with different size and will be extended with double-mirror 
Schwarzschild-Coude telescopes to further improve angular resolution and sensitivity over wider field of view.

One of the key component of these instruments is the imaging camera that usually comprises a large number of photosensors ($\sim$1000 photomultiplier tubes, or possibly even $\sim$10000 silicon photosensors 
for wide field of view telescopes). The Cherenkov light from extensive air showers flashes the camera, resulting in the images of certain shape and intensity distribution. 

However, telescopes are operated under conditions of relatively bright night sky background light and the imaging camera data always contain these undesired noise signals with rate values reaching up to 1GHz per camera pixel
during moon-time observations. 
Also, photosensor intrinsic noise pulses can affect the quality of shower images. Therefore camera image cleaning procedure is clearly necessary, prior to further analysis steps.


During the last decade, the implementation of fast telescope readout systems based on Flash ADC samplers has allowed the reconstruction of the signal arrival time in every photosensor of Cherenkov images. The individual waveform sampling also offered wide scope for signal extraction and noise reduction optimization.

\section{Image Cleaning Method}

The ultimate task of an image cleaning procedure is to determine the maximal amount of pixels with the signal from the shower, keeping to a minimum the number of pixels with noise signals.
Thus, a maximum of information initially contained in the data will be provided for further analysis.

One should keep in mind that larger numbers of reconstructed pixels do not always result in better performance of the gamma/hadron separation, 
since usually analysis methods are strongly optimized to traditional image processing and signal extraction methods. Therefore, to realize the full potential of more 
complete shower images analysis methods should be widely revised and adopted, but we would like to keep these issues outside of the scope of the present paper.

In the traditional image cleaning method pixels with background signals are rejected by applying a charge cut (core threshold) on the pixel signal. Those  pixels which were not rejected by the charge cut are assigned to be
an image core candidate. Next, the topological condition of having at least one neighbor is applied and image core candidates without a neighbor are discarded. In the last step,  the vicinity of the selected core 
pixels is revised again and the lower charge cut (boundary threshold) is applied to complete the image with boundary pixels. The necessary threshold values scale with the root mean square of the background fluctuations. Here one should note that the short time duration and fine time structure of shower images is not taken into account, hence all noise signals induced within the readout window of the data acquisition system are contributing to the background.

 A natural improvement of the traditional image cleaning method can be achieved by considering the additional image time structure information, so that for neighboring pixels not only signal amplitudes, but also arrival time differences
are examined. In this fashion, the unnecessary large phase space of accepted noise can be significantly reduced and more closely match to the shower image phase space. Moreover, the next-neighbor groups of pixels with multiplicities 
larger then two can be considered as units for image formation. These ideas were successfully implemented in the novel next-neighbor image cleaning procedure \cite{bib:shayduk}, developed for  the
MAGIC telescopes with a relatively fast readout system. The traditional image cleaning algorithms were also revised earlier by introducing  the wavelet analysis \cite{bib:wavelet},  but important image timing information was not included or available in that studies.  

 \begin{figure}[t]
  \centering
  \includegraphics[width=0.4\textwidth]{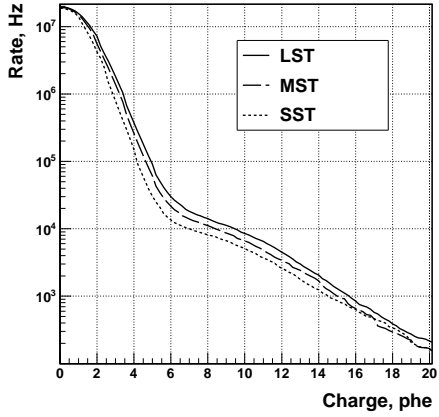}
  \caption{Differential charge spectra in a single pixel for three types of CTA telescopes under exposure to night sky background light of typical intensities for extra-galactic observations (Monte-Carlo simulations). 
The charge was integrated over 4 ns time slice and converted to photoelectrons  (phe). LST, MST and SST acronyms stand for Large-, Medium- and Small- Size Telescope accordingly. The transition region from noise charges, induced by light of the night sky to the regime, dominated by photosensor after-pulses is revealed at $\sim$6 phes. The rate saturation for charges below 2 phes is due to the 50 ns gate-width of the counter.}
  \label{ipr_fig}
 \end{figure}

In the image cleaning procedure, described in \cite{bib:shayduk}, the image is formed by the next-neighbor groups of pixels  with different multiplicity  (2-nn, 3-nn and 4-nn groups). In order to assign such a group to the shower image, pixels in the
 group are required to have the charge above a certain threshold and pulses time coincidences should be below a certain value.  The required thresholds and coincidence time limits are defined by the background fluctuations and the
group multiplicity. In this notation, the next-neighbor group search with the multiplicity of two ( and  the time coincidence condition loosened to include the full readout window) corresponds to the core pixel search in the traditional image cleaning method, described above. 

However, in the next-neighbor image cleaning procedure  currently used, threshold values and coincidence time limits
are static orthogonal cuts in the phase plane, formed by minimal pixel charge $Q$ in the group and maximal time difference $\Delta T$.  

Meanwhile, the noise differential probability contour in this $Q$ - $\Delta T$  phase plane is obviously a smooth function of threshold and coincidence time. This issue has  become important with the demanding  requirements of CTA, especially considering the much larger energy range and camera field of view, 
the low light content images could have significantly broader time spread, compared to typical images of low intensity in the MAGIC telescope data.  

The natural way to adopt the next-neighbor cleaning algorithm to the extended
phase space of shower images is to use a dynamical cut in the group $Q$ - $\Delta T$ parameter plane. The shape of this dynamical cut curve can be obtained from noise properties of the individual pixel, assuming that 
all pixels in the camera are the same in this term.

 \begin{figure}[t]
  \centering
  \includegraphics[width=0.42\textwidth]{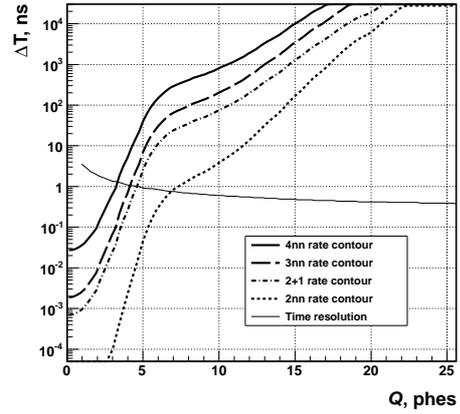}
  \caption{
  Parameter plane, formed by minimal charge $Q$ and maximal coincidence time $\Delta$T in the tested next-neighbor group. Contours of constant rate are shown for rate value of $\sim3$kHz, which corresponds to less then 1\% probability of the fake image if the shower event recorded over 50 ns time window. The pattern of 2-nn group and one second-nearest neighbor is denoted as 2+1 group. The noise region of higher accidental rates of the group is above the corresponding curve.  For the 2-nn group, the after-pulsing noise of the photosensor becomes significant in the strongly relevant for shower image reconstruction  $\Delta T$  range of 1-10 ns.  }
  \label{contour_fig}
 \end{figure}

The typical differential noise rate in one pixel  is presented in Fig. \ref{ipr_fig}. 
Noise charges higher then $\sim$ 6 photoelectrons are induced mainly by the intrinsic noise of the photosensor - after-pulsing effect, described relevantly to CTA in \cite{bib:ap}.  
The differential accidental rate $\Re_{acc}$ of the next-neighbor groups in the camera with multiplicity $n$ can be derived from the individual pixel noise rate $R_{pix}(Q)$ as following:

\begin{equation}
\Re_{acc}(n,Q,\Delta T)= C_{n} \cdot \Delta T^{n-1} \cdot R_{pix}(Q)^{n},
\end{equation}

where $Q$ - is the charge in one pixel, $\Delta T$ - is the time coincidence window and $C_{n}$ - combinatorial factor, depending on the group multiplicity and total number of the pixels in the camera. 
The contour for the constant differential rate $\Re_{acc}=const$ is described by the following equation for the time coincidence $\Delta T$:

\begin{equation}
\Delta T(\Re_{acc}[Hz], Q)= \exp[ \frac{1}{n-1} \cdot \ln(\frac{\Re_{acc}[Hz]}{C_{n} \cdot R_{pix}(Q)^{n}}) ]
\label{eq_contour}
\end{equation}

This contour can serve as a dynamic cut in the $Q-\Delta T$ parameter plane, so that any next-neighbor group with parameter values, above this contour should  be rejected as a background group.
The value of the differential rate $\Re_{acc}[Hz]$ can be chosen, according to the desired probability level of accidental images. 
Contours of constant accidental rate for next-neighbor  groups, calculated with equation \ref{eq_contour}  are shown in Fig.\ref{contour_fig}. Thresholds for core and boundary pixel search from the traditional image cleaning method would appear in the plot  as vertical lines at positions of 10 and 5 phes correspondingly.  The 2-nn group threshold curve shape is fully determined by the photosensor after-pulsing additional noise
for coincidence times above $\sim1$ ns.  For further tuning of of the method to the fine shower image features, groups with higher multiplicities or even certain patterns can be easily added to the search algorithm.

 \begin{figure}[t]
  \centering
  \includegraphics[width=0.45\textwidth]{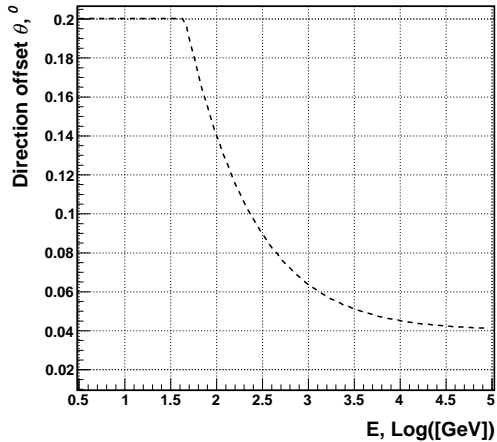}
  \caption{Reconstructed  direction offset $\theta=|\theta_{true}-\theta_{rec}|$ versus energy. The curve represents the dynamic cut, applied in the analysis of $\gamma$-ray events and corresponds to 
  the typical angular resolution for arrays like H.E.S.S and VERITAS. }
  \label{thetacut_fig}
 \end{figure}
 
  \begin{figure}[t]
  \centering
  \includegraphics[width=0.45\textwidth]{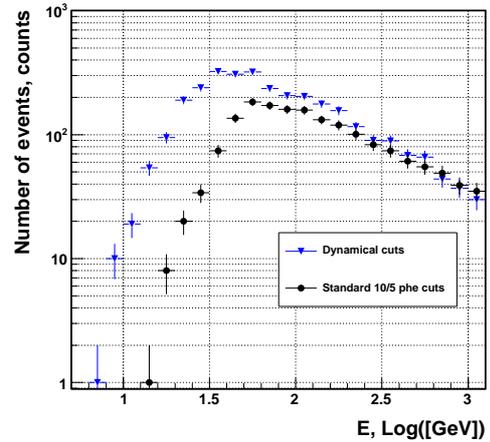}
  \caption{Number of events with well-reconstructed direction versus $\gamma$-ray energy for LST telescopes.  Monte-Carlo $\gamma$-ray events were simulated with power law  $dN/dE \sim E^{-\alpha}$, $\alpha=2$. The data was processed with traditional image cleaning with thresholds of 10 and 5 phes  for core and boundary pixels accordingly (circles) and the new image cleaning, described in this paper (triangles). The gain in number of events for optimized next-neighbor image cleaning algorithm is  very prominent for energies below 100 GeV.}
  \label{icgain_fig}
 \end{figure}
\section{Results}

The performance of the optimized next-neighbor image cleaning method was studied with  Monte-Carlo data for the CTA  experiment.
The $\gamma$-ray extensive air  showers were simulated  with a power law energy spectrum $dN/dE \sim E^{-2}$, using the \emph{SimTel} package \cite{bib:simtel}.  
The telescope triggers were simulated in great detail  by the trigger simulation code \emph{TrigSim} \cite{bib:trigsim}.
 
The photoelectrons, induced by showers in imaging cameras
were converted to Flash ADC traces and the corresponding background light of the night sky with  intrinsic noise of photosensors were added to the camera data. From  these simulated Flash ADC traces the 
charge and the arrival time in every camera pixel were extracted and the image cleaning procedure was applied. The shower direction reconstruction was performed with the VERITAS  analysis code \cite{bib:evndisp}.  Since events with poor/failed direction reconstruction are usually useless for next analysis steps they were discarded by the dynamic direction offset cut, presented in Fig.\ref{thetacut_fig}.  
Then  the number of well-reconstructed events in each energy bin was compared  for two image cleaning methods: traditional 10/5 phes  and the optimized next-neighbor image cleaning.  The result of this comparison for CTA Large Size Telescopes is shown in Fig.\ref{icgain_fig}, revealing the remarkable gain for the new method  in the energy range of 10-100 GeV.

\section{Conclusions}

The suggested image cleaning method exploits the time structure of the shower flash, but does not constrain the total time spread of the image, can be applied to events within a wide energy range and keeps more image features, providing room for novel sophisticated $\gamma$/hadron separation methods. 

\vspace*{0.5cm}
\footnotesize{{\bf Acknowledgment:}{The ICRC 2013 is funded by FAPERJ, CNPq, FAPESP, CAPES and IUPAP.}}

\end{document}